\documentstyle[11pt,epsf]{article}

\begin{document}
\hfill hep-th/0507147\\
 \begin{center}
{\Large \bf Flow Equations for Uplifting Half-Flat to Spin(7)
Manifolds} \vskip 0.1in \Large
Aalok Misra\footnote{e-mail: aalokfph@iitr.ernet.in}\\
Department of Physics, Indian Institute of Technology Roorkee,
Roorkee 247 667, India\\
and\\
Jefferson Physical Laboratory, Harvard University, Cambridge, MA
02138,
USA\\
\vspace{0.5 cm}
\end{center}

\begin{abstract}
In this supplement to \cite{fkmr}, we discuss the uplift of
half-flat six-folds to Spin(7) eight-folds by fibration of the
former over a product of two intervals. We show that the same can be
done in two ways - one, such that the required Spin(7) eight-fold is
a double $G_2$ seven-fold fibration over an interval, the $G_2$
seven-fold itself being the half-flat six-fold fibered over the
other interval, and second, by simply considering the fibration of
the half-flat six-fold over a product of two intervals. The flow
equations one gets are an obvious  generalization of the Hitchin's
flow equations (to obtain seven-folds of $G_2$ holonomy from
half-flat six-folds \cite{hitchin}). We explicitly show the uplift
of the Iwasawa using both methods, thereby proposing the form of new
Spin(7) metrics. We give a plausibility argument ruling out the
uplift of the Iwasawa manifold to a Spin(7) eight fold at the
``edge", using the second method. For $Spin(7)$ eight-folds of the
type $X_7\times S^1$, $X_7$ being a seven-fold of $SU(3)$ structure,
we motivate the possibility of including elliptic functions into the
``shape deformation" functions of seven-folds of $SU(3)$ structure
of \cite{fkmr} via some connections between elliptic functions, the
Heisenberg group, theta functions, the already known $D7$-brane
metric\cite{greenetal} and hyper-K\"{a}hler metrics obtained in
twistor spaces by deformations of Atiyah-Hitchin manifolds by a
Legendre transform in \cite{c}.
\end{abstract}
\newpage

It is known that manifolds with $G_2$ and Spin(7) holonomies, are
very useful in getting minimal amount of supersymmetry after
compactification of seven and eight dimensions, respectively, in
string/M-theory \cite{Gubseretal}. In the past few years, half-flat
manifolds have been shown to be relevant to flux compactifications
in string theory (see \cite{cardosoetal,gurierietal} and references
therein). In \cite{fkmr}, using the results of \cite{dp}, we
explicitly showed how to uplift the Iwasawa manifold, an example of
a half-flat manifold, to seven-folds of either $G_2$-holonomy or
$SU(3)$ structure. In this short note,  we show how to uplift
half-flat manifolds to Spin(7) eight-folds. We will be following the
notations of \cite{fkmr}. We also give plausibility arguments in
favor of inclusion of the Weierstrass elliptic function in
$Spin(7)$-metrics of the type $X_7\times S^1$, $X_7$ being
seven-folds of $SU(3)$ structure.

Spin(7) folds are characterized by a self-dual closed (and hence
co-closed) Cayley four-form (with the additional constraint that the
$\hat{A}$-genus of the eight-fold equals unity \cite{joyce}). We
will begin with the construction of a Spin(7) eight-fold as a double
$G_2$-fibration over an interval. We then go on to constructing a
Spin(7) eight-fold as a fibration of a half-flat over the product of
two intervals.

\section*{Spin(7) as a Double $G_2$-fibration over an interval}

Let us begin with the following construction of a Spin(7) eight-fold
$X_8^{\rm Spin(7)}$:
\begin{eqnarray*}
\label{eq:G2fibr}
& & \begin{array}{clll}I_{t_1} & & &\\
\uparrow & & \\
\ \vert\ &\!\!\!\!\!\!\!\! X^{G_2}_7(t_2)&\!\!\!\!\!\!\!\!
\longrightarrow & \!\!\!\!\!I_{t_2} \\
\vert & & \!\!\!\!\!\!\! M_6 & \\
\vert & & & \\
X_8^{\rm Spin(7)} & \longrightarrow & \!\!\!\!\!\!\!\!I_{t_2}& \\
& X_7^{G_2}(t_1) & & \\
& \ \ \ | & & \\
& \ \ \ \!\downarrow
M_6 & &  \\
& \ \ I_{t_1} & & \\
\end{array}\nonumber\\
& & \hskip 3cm Fig. 1
\end{eqnarray*}
In other words, one has a ``double" $G_2$-fibration structure in the
sense that the Spin(7) eight-fold is a fibration of a
$G_2(t_1)$-manifold over an interval $I(t_2)$ {\it as well as} a
$G_2(t_2)$-manifold over an interval $I(t_1)$, where the
$G_2(t_i)$-manifold is itself a fibration of a half-flat six-fold
over an interval $I(t_i)$ (via the Hitchin's construction
\cite{hitchin}). Given a half-flat manifold $M_6(\Omega,J)$, the
Hitchin's construction involves (w.r.t. seven dimensions) a closed
and co-closed three-form $\phi_3=J\wedge dt + \Omega_+(t)$,
$\Omega_+\equiv{\rm Re}(\Omega)$, one can write down the following
Cayley four-form $\phi_4$:
\begin{eqnarray}
\label{eq:correct4form} & &
\phi_4=\alpha_1*_7\phi_3^1+\alpha_2*_7\phi^2_3+\tilde{\alpha}_1\phi^1
\wedge dt_2+\tilde{\alpha}_2\phi^2_3\wedge dt_1\nonumber\\
& & + \beta_1\Omega_+\wedge dt_1 + \beta_2\Omega_+\wedge dt_2 +
\gamma_1\Omega_-\wedge dt_1 + \gamma_2\Omega_-\wedge dt_2
+ \delta J\wedge J + \epsilon J\wedge dt_1\wedge dt_2.\nonumber\\
& &
\end{eqnarray}
The self-duality condition would thus imply:
\begin{eqnarray}
\label{eq:selfdual} & & \alpha_1\biggl(J\wedge dt_1\wedge dt_2 +
\Omega_+\wedge dt_2\biggr)
+\alpha_2\biggl(J\wedge dt_1\wedge dt_2 - \Omega_+\wedge dt_1\biggr)\nonumber\\
& & +\tilde{\alpha}_1\biggl({1\over2}J\wedge J + \Omega_-\wedge
dt_1\biggr) +\tilde{\alpha}_2\biggl(-{1\over2}J\wedge J -
\Omega_-\wedge dt_2\biggr)
\nonumber\\
& & -\beta_1\Omega_-\wedge dt_2+\beta_2\Omega_-\wedge
dt_1+\gamma_1\Omega_+
\wedge dt_2 - \gamma_2\Omega_+\wedge dt_1\nonumber\\
& & +2\delta J\wedge dt_1\wedge dt_2 + {\epsilon\over2}J\wedge J\nonumber\\
& & = \alpha_1\biggl({1\over2}J\wedge J + \Omega_-\wedge dt_1\biggr)
+
\alpha_2\biggl({1\over2}J\wedge J + \Omega_-\wedge dt_2\biggr)\nonumber\\
& & +\tilde{\alpha}_1\biggl(J\wedge dt_1\wedge dt_2 + \Omega_+\wedge
dt_2) + \tilde{\alpha}_2(-J\wedge dt_1\wedge dt_2 + \Omega_+\wedge
dt_1
\biggr)\nonumber\\
& & +\beta_1\Omega_+\wedge dt_1+\beta_2\Omega_+\wedge
dt_2+\gamma_1\Omega_-\wedge
dt_1+\gamma_2\Omega_-\wedge dt_2\nonumber\\
& & \delta J\wedge J + \epsilon J \wedge dt_1\wedge dt_2,
\end{eqnarray}
implying:
\begin{eqnarray}
& &
\alpha_1+\alpha_2+2\delta=\tilde{\alpha}_1-\tilde{\alpha}_2+\epsilon,
\nonumber\\
& & \alpha_1+\gamma_1=\tilde{\alpha}_1+\beta_2,\ -\alpha_2-\gamma_2
=\tilde{\alpha}_2+\beta_1.
\end{eqnarray}
Hence, the following is the form of the self-dual four-form:
\begin{eqnarray}
\label{eq:selfdualphi4} & &
\phi_4=\alpha_1*_7\phi_3^1+\alpha_2*_7\phi^2_3+\tilde{\alpha}_1\phi^1\wedge
dt_2+\tilde{\alpha_2}\phi_3^2\wedge dt_1\nonumber\\
& & +\beta_1\Omega_+\wedge dt_1+\beta_2\Omega_+\wedge
dt_2+(\tilde{\alpha}_1
+\beta_2-\alpha_1)\Omega_-\wedge dt_1\nonumber\\
& & -(\alpha2+\tilde{\alpha}_2+\beta_1)\Omega_-\wedge dt_2 + \delta
J\wedge J
+(\alpha_1+\alpha_2+2\delta+\tilde{\alpha}_2-\tilde{\alpha}_1)J\wedge
dt_1\wedge
dt_2.\nonumber\\
& &
\end{eqnarray}
For (\ref{eq:selfdualphi4}) to be a Cayley four-form, it must
satisfy the condition that it is closed, which on using:
\begin{eqnarray}
\label{eq:G2} & & d*_7\phi^1_3=(\hat{d}+dt_1\wedge
{\partial\over\partial t_1} + dt_2\wedge {\partial\over\partial
t_2})*_7\phi^1_3=dt_2\wedge {\partial*_7\phi_3^1
\over\partial t_2}\nonumber\\
& & d*_7\phi^2_3=(\hat{d}+dt_1\wedge {\partial\over\partial t_1} +
dt_2\wedge {\partial\over\partial t_2})*_7\phi^2_3=dt_1\wedge
{\partial*_7\phi_3^2
\over\partial t_2}\nonumber\\
& & d\phi_3^1=dt_2\wedge{\partial\phi_3^1\over\partial t_2},\
d\phi_3^2=dt_2\wedge{\partial\phi_3^2\over\partial t_1},
\end{eqnarray}
implies:
\begin{eqnarray}
\label{eq:dphi40} & & d\phi_4|_{\hat{d}J\wedge J=\hat{d}\Omega_+=0}
= \alpha_1\biggl(dt_2\wedge{\partial J\over\partial t_2}\wedge J +
dt_2\wedge{\partial\Omega_-\over\partial t_2}\wedge dt_1\biggr)
\nonumber\\
& & +\alpha_2\biggl(dt_1\wedge{\partial J\over\partial t_1}\wedge J
+ dt_1\wedge {\partial\Omega_-\over\partial t_1}\wedge dt_1\biggr)
+\beta_1(dt_2\wedge{\partial\Omega_+\over\partial t_2}\wedge dt_1)
+\beta_2(dt_1\wedge{\partial\Omega_+\over\partial t_1}\wedge dt_2)
\nonumber\\
& & +(\tilde{\alpha}_1+\beta_2-\alpha_1)
\biggl(\hat{d}\Omega_1\wedge dt_1
+dt_2\wedge{\partial\Omega_-\over\partial t_2}\wedge dt_1
\biggr)\nonumber\\
& & -(\alpha_2+\tilde{\alpha}_2+\beta_1)
\biggl(\hat{d}\Omega_1\wedge dt_2
+dt_1\wedge{\partial\Omega_-\over\partial t_1}\wedge dt_2\biggr)
+2\delta\biggl(dt_1\wedge{\partial J\over\partial t_1}\wedge J +
dt_2\wedge
{\partial J\over\partial t_2}\wedge J\biggr)\nonumber\\
& & +(\alpha_1+\alpha_2+2\delta+\tilde{\alpha}_2-\tilde{\alpha}_1)
\hat{d}J\wedge dt_1\wedge dt_2=0.
\end{eqnarray}
One thus gets the following flow equations:
\begin{eqnarray}
\label{eq:flowSpin7G2} & & (\alpha_1+2\delta){\partial
J\over\partial t_2}\wedge J =
(\alpha_2+\tilde{\alpha}_2+\beta_1)\hat{d}\Omega_-,\nonumber\\
& & (2\delta + \alpha_2){\partial J\over\partial t_1}\wedge J
=(\alpha_1-\beta_2+\tilde{\alpha}_1)\hat{d}\Omega_-,\nonumber\\
& & (\tilde{\alpha}_1+\tilde{\beta}_2){\partial\Omega_-\over\partial
t_2} +(\tilde{\alpha}_2+\beta_1){\partial\Omega_-\over\partial t_1}
+\beta_1{\partial\Omega_+\over\partial
t_2}-\beta_2{\partial\Omega_+\over\partial t_1}
+(\alpha_1+\alpha_2+2\delta+\tilde{\alpha}_2-\tilde{\alpha}_1)\hat{d}J=0.
\nonumber\\
& &
\end{eqnarray}
Let us use the flow equations of (\ref{eq:flowSpin7G2}) to
explicitly uplift the Iwasawa manifold to a Spin(7) eight-fold,
working with the standard complex structure limit of the Iwasawa,
and consider its deformation of the type:
\begin{equation}
\label{eq:Jdef}
J(t_1,t_2)=e^{a(t_1,t_2)}e^{12}+e^{b(t_1,t_2)}e^{34}+e^{c(t_1,t_2)}e^{56},
\end{equation}
and
\begin{equation}
\label{eq:Omegadef}
\Omega(t_1,t_2)=e^{{(a+b+c)(t_1,t_2)\over2}}(e^1+ie^2)\wedge(e^3+ie^4)\wedge
(e^5+ie^6).
\end{equation}
Then the flow equations (\ref{eq:flowSpin7G2}) imply:
\begin{itemize}
\item
\begin{equation}
\label{eq:sol1} {\partial a\over\partial t_1}={\partial
b\over\partial t_1}=-{\partial c \over\partial t_1},\ 2{\partial
a\over\partial t_1}e^{a+b}=4\xi_1 e^{{a+b+c \over2}},
\end{equation}
where $\xi_1\equiv{\alpha_1-\beta_2+\tilde{\alpha}_1\over
2\delta+\alpha_2}$, which could be satisfied by equality of
$a,b,-c$, and:
\begin{equation}
\label{eq:sol2}
a(t_1,t_2)={2\over3}ln\biggl(3\xi_1e^{\lambda_1}t_1+f_2(t_2)\biggr),
\end{equation}
$\lambda_1$ being a linear combination of the integration constants
that would appear in the integration of the first set of equations
in (\ref{eq:sol1});
\item
similarly,
\begin{equation}
\label{eq:sol3}
a(t_1,t_2)={2\over3}ln\biggl(3\xi_2e^{\lambda_1}t_2+f_1(t_1)\biggr),
\end{equation}
where
$\xi_2\equiv{\alpha_2+\tilde{\alpha_2}+\beta_1\over\alpha_1+2\delta}$.
\item
\begin{eqnarray}
\label{eq:sol4} & & {(\tilde{\alpha}_1+\beta_2)\over2}{\partial
a\over\partial t_2}
+{(\tilde{\alpha}_2+\beta_1)\over2}{\partial a\over\partial t_1}=0,\nonumber\\
& & \Leftrightarrow
(\tilde{\alpha}_1+\beta)\xi_2+(\tilde{\alpha}_2+\beta_1)
\xi_1=0,\nonumber\\
& & e^{a+b+c\over2} \biggl({\beta_1\over2}{\partial a\over\partial
t_2} - {\beta_2\over2}{\partial a \over\partial t_1}\biggr) =
(\alpha_1+\alpha_2+2\delta+\tilde{\alpha}_2-\tilde{\alpha}_1)e^c,
\nonumber\\
& & \Leftrightarrow
(\beta_1\xi_2-\beta_2\xi_1)e^{-{3a\over2}+\lambda_1} =
(\alpha_1+\alpha_2+2\delta+\tilde{\alpha}_2-\tilde{\alpha}_1)e^{-a+\lambda_3},
\nonumber\\
& & \Rightarrow \beta_1\xi_2-\beta_2\xi_1 =
\alpha_1+\alpha_2+2\delta+\tilde{\alpha}_2-\tilde{\alpha}_1.
\end{eqnarray}
\end{itemize}

Hence, the metric corresponding to the Spin(7) eight-fold obtained
by uplifting the Iwasawa manifold via the flow equations of
(\ref{eq:flowSpin7G2}) such that the eight-fold is a double
$G_2$-fibration over an interval, is given by the following
solutions to (\ref{eq:sol1}) to (\ref{eq:sol4}):
\begin{equation}
\label{eq:Spin7metric} ds_8^2=ds^2_{I\times I}(t_1,t_2)+(1+\xi_1 t_1
+ \xi_2 t_2)^{{2\over 3}}(|dz|^2+|dv|^2) +{1\over(1 + \xi_2 t_1 +
\xi_2 t_2)^{{2\over3}}}|du-z dv|^2,
\end{equation}
with the constraints:
\begin{eqnarray}
\label{eq:constraints} & & \beta_1\xi_2-\beta_2\xi_1 =
\alpha_1+\alpha_2+2\delta +\tilde{\alpha}_2-
\tilde{\alpha}_1,\nonumber\\
& &
{\xi_1\over\xi_2}=-{\tilde{\alpha}_1+\beta_2\over\tilde{\alpha}_2+\beta_1},
\ {\rm or}\ \tilde{\alpha}_1=-\beta_2,\ \tilde{\alpha}_2=-\beta_1.
\end{eqnarray}
Notice that (\ref{eq:Spin7metric}) has the required double
$G_2$-fibration structure of Fig.1 by noting that
(\ref{eq:Spin7metric}) is made up of:
\begin{equation}
\label{eq:G21} ds_7^2(t_1,{\rm given}\ t_2)= dt_1^2 +
(\xi^{\prime}_1 + \xi_1 t_1)^{{2\over3}}(|dz|^2+|dv|^2)
+{1\over(\xi_1^\prime+\xi_1 t_1)^{{2\over3}}}|du-z dv|^2,
\end{equation}
and
\begin{equation}
\label{eq:G22} ds_7^2(t_2,{\rm given}\ t_1) = dt_2^2 + (\xi^\prime_2
+ \xi_2 t_2)^{{2\over3}}(|dz|^2+|dv|^2) +{1\over(\xi_2^\prime+\xi_2
t_2)^{{2\over3}}}|du-z dv|^2,
\end{equation}
which are the two-parameter $G_2$-metrics of \cite{fkmr}
\footnote{In \cite{fkmr}, however, one had set $\xi_i^\prime=1$.}
The metric of (\ref{eq:Spin7metric}) also thus has
$G_2$-holonomy\footnote{We thank J.Maldacena for pointing this
out.}.

\section*{Spin(7) as a fibration of a Half-flat over $I\times I$}

Let us now consider the following fibration structure:
\begin{equation}
\begin{array}{ccc}
X^{\rm Spin(7)}_8 & \longrightarrow & I_{t_1}\times I_{t_2}\\
& M_6 & \\
\end{array}
\end{equation}
Let us assume that the Cayley four-form is given by:
\begin{equation}
\phi_4=a_1\Omega_-\wedge dt_1 + a_2\Omega_-\wedge dt_2 +
b_1\Omega_+\wedge dt_1 +b_2\Omega_+\wedge dt_2 + cJ\wedge J + f
J\wedge dt_1\wedge dt_2.
\end{equation}
One hence gets:
\begin{equation}
*_8\phi_4=a_1\Omega_+\wedge dt_2 - a_2\Omega_+\wedge dt_1 - b_1\Omega_-\wedge
dt_2+b_2\Omega_-\wedge dt_1+2cJ\wedge dt_1\wedge dt_2 +
{f\over2}J\wedge J,
\end{equation}
implying that for $\phi_4=*_8\phi_4$,
\begin{equation}
a_1=b_1,\ a_2=-b_1,\ c={d\over2}.
\end{equation}
The required Cayley four-form is:
\begin{equation}
\label{eq:Cayley} \phi_4=a_1\Omega_-\wedge dt_1+a_2\Omega_-\wedge
dt_2 -a_2\Omega_+\wedge dt_1+a_1\Omega_+ \wedge
dt_2+{f\over2}J\wedge J + f J\wedge dt_1\wedge dt_2.
\end{equation}
Finally, the condition that $\phi_4$ of (\ref{eq:Cayley}) is closed
gives:
\begin{eqnarray}
& & d\phi_4=a_1\biggl(\hat{d}\Omega_-\wedge dt_1
+dt_2\wedge{\partial\Omega_-\over\partial t_2} \wedge
dt_1\biggr)+a_2\biggl( \hat{d}\Omega_-\wedge dt_2 +
dt_1\wedge{\partial\Omega_-\over
\partial t_1}\wedge dt_2\biggr)\nonumber\\
& & +a_1\biggl( \hat{d}\Omega_+\wedge
dt_2+dt_1\wedge{\partial\Omega_+\over\partial t_1} \wedge
dt_2\biggr) - a_2\biggl(\hat{d}\Omega_+\wedge dt_1 + dt_2\wedge
{\partial
\Omega_-\over\partial t_2}\wedge dt_1\biggr)\nonumber\\
& & + f\biggl(\hat{d}J\wedge J +dt_1\wedge{\partial J\over\partial
t_1}\wedge J + dt_2\wedge{\partial J \over\partial t_2}\wedge
J\biggr) = 0.
\end{eqnarray}
Using that for half-flat manifolds, $\hat{d}J\wedge
J=\hat{d}\Omega_+=0$, one thus gets the following flow equations:
\begin{eqnarray}
\label{eq:floweqns}
& &a_1\ \hat{d}\Omega_- = - f\ {\partial J\over\partial t_1}\wedge J,\nonumber\\
& & a_2\ \hat{d}\Omega_- = -f\ {\partial J\over\partial t_2}\wedge J,\nonumber\\
& & -a_1\ {\partial\Omega_+\over\partial t_2} + a_2\
{\partial\Omega_-\over\partial t_1} + a_2\
{\partial\Omega_+\over\partial t_2} + a_1\
{\partial\Omega_+\over\partial t_2}= f\ \hat{d}J.
\end{eqnarray}

One can again show that one can explicitly uplift the Iwasawa
manifold to a Spin(7) eight-fold at standard complex structure limit
of the Iwasawa and consider its deformation of the type as given in
(\ref{eq:Jdef}) and (\ref{eq:Omegadef}). The set of equations that
one gets from (\ref{eq:floweqns}), are:
\begin{eqnarray}
\label{eq:solmethod2} & & {\partial a\over\partial t_i}={\partial
b\over\partial t_i}=-
{\partial c\over\partial t_i},\nonumber\\
& & {\partial a\over\partial t_1}=-{2a_1\over
f}e^{-{3a\over2}+\lambda_1},\ {\partial a\over\partial
t_2}=-{2a_2\over f}e^{-{3a\over2}+\lambda_1},
\nonumber\\
& & \biggl( {a_1\over2}{\partial a\over\partial t_2}-a_2{\partial
a\over\partial t_1}\biggr) e^{{a+b+c\over2}}=0, \
\biggl(-{a_2\over2}{\partial a\over\partial t_2} - {a_1\over2}
{\partial a\over\partial t_1}\biggr)e^{{a+b+c\over2}} =f e^c,
\end{eqnarray}
which are satisfied by:
\begin{eqnarray}
\label{eq:sol} & & a(t_1,t_2) = {2\over3} ln\biggl(1 +
3{a_1e^{\lambda_1}\over f}t_1 + 3{a_2e^{\lambda_2} \over
f}t_2\biggr),
\nonumber\\
& & {\rm with}\ a_1^2+a_2^2=f^2.
\end{eqnarray}
Thus, the metric for the Spin(7) eight-fold is:
\begin{eqnarray}
\label{eq:hfxintxint} & & ds_8^2 = ds^2_{I\times I}(t_1,t_2) +
\biggl(1 + {a_1\over\sqrt{a_1^2+a_2^2}}t_1 +
{a_2\over\sqrt{a_1^2+a_2^2}} t_2\biggr)^{{2\over3}}(|dz|^2 + |dv|^2)
\nonumber\\
& & + \biggl(1 + {a_1\over\sqrt{a_1^2+a_2^2}}t_1 +
{a_2\over\sqrt{a_1^2+a_2^2}} t_2\biggr)^{-{2\over3}}|du - v dz|^2.
\end{eqnarray}
Again, the metric also has $G_2$-holonomy.

However, it is unlikely to be able to uplift the Iwasawa to a
Spin(7) eight-fold at the ``edge". One notes that at the edge (See
\cite{cardosoetal} and references therein), the one-forms,
incorporating $t_{1,2}$-dependent deformations, are:
$\alpha=-e^{{\cal A}(t_1,t_2)}f^1,\beta=e^{{\cal
B}(t_1,t_2)}(f^3+if^4), \gamma=e^{{\cal C}(t_1,t_2)}(e^5+ie^6)$, and
$J={i\over2}(\alpha\wedge
{\bar\alpha}+\beta\wedge{\bar\beta}+\gamma\wedge{\bar\gamma}$ and
$\Omega=\alpha\wedge\beta\wedge\gamma)$. The one-forms $f^i,
i=1,...,4$ are defined via $f^i=P^i_j e^j$, where $P\in SO(4)$
matrix, and one write it as $\left(\begin{array}{cc}
X & 0\\
0 & Y \\
\end{array}\right)$, where $X,Y\in SU(2)$, i.e.,
$\left(\begin{array}{cccc}
P^1_1 & P^1_2 & 0 & 0 \\
P^2_1 & P^2_2 & 0 & 0 \\
0 & 0 & P^3_3 & P^3_4 \\
0 & 0 & P^4_3 & P^4_4 \\
\end{array}\right)$, where $P^1_1P^2_2 - P^1_2P^2_1=P^3_3P^4_4-P^3_4P^4_3=1$.
The flow equations $\hat{d}\Omega_-=-{\partial J\over\partial
t_1}\wedge J=-{\partial J \over\partial t_2}\wedge J$ implies ${\cal
A}$ gives the same result after differentiation w.r.t. $t_1$ or
$t_2$. Unlike the standard complex structure limit, there are common
components to $\Omega_+$ and $\Omega_-$ in the edge. One can show
that the other flow equation $a_1{\partial\Omega_-\over\partial t_2}
-a_2{\partial\Omega_-\over\partial t_1} -
a_2{\partial\Omega_+\over\partial t_2} -
a_1{\partial\Omega_+\over\partial t_1} + f \hat{d}J=0$ becomes:
\begin{eqnarray}
\label{eq:flow2edge} & & e^{{\cal A}+{\cal B}+{\cal C}}\Biggl[
\biggl( {\cal D}_- (P^1_1P^3_3+P^2_1P^4_3) - {\cal D}_+
(P^2_1P^3_3-P^1_1P^4_3)\biggr)e^{136}\nonumber\\
& & + \biggl( {\cal D}_-(P^1_1P^3_4+P^2_1P^4_4)-
{\cal D}_+(P^2_1P^3_4-P^1_1P^4_4)\biggr)e^{146}\nonumber\\
& & + \biggl( {\cal D}_-(P^1_2P^3_3+P^2_2P^4_3) -
{\cal D}_+(P^2_2P^3_3-P^1_2P^4_3)\biggr)e^{236}\nonumber\\
& & + \biggl( {\cal D}_-(P^1_2P^3_4+P^2_2P^4_4 -
{\cal D}_+(P^2_2P^3_4-P^1_2P^4_4)\biggr)e^{246}\nonumber\\
& & + \biggl( {\cal D}_-(-P^1_1P^4_3+P^2_1P^3_3)-
{\cal D}_+(-P^1_1P^3_3-P^2_1P^4_3)\biggr)e^{135}\nonumber\\
& & + \biggl( {\cal D}_-(-P^1_1P^4_4+P^2_1P^3_4) -
{\cal D}_+(-P^1_1P^3_4-P^2_1P^4_4)\biggr)e^{145}\nonumber\\
& & + \biggl( {\cal D}_-(-P^1_2P^4_3+P62_2P^3_3) -
{\cal D}_+(-P^1_2P^3_3-P^2_2P^4_3)\biggr)e^{235}\nonumber\\
& & + \biggl( {\cal D}_-(-P^1_2P^4_4+P^2_2P^3_4) - {\cal
D}_+(-P^1_2P^3_4-P^2_2P^4_4)\biggr)e^{246}\Biggr]
\nonumber\\
& & =-e^{2{\cal C}}\biggl(e^{135}+e^{425}-e^{614}-e^{623}\biggr),
\end{eqnarray}
(where ${\cal D}_+\equiv \biggl(a_2{\partial\over\partial t_2} +
a_1{\partial\over\partial t_1}\biggr){\cal A}{\cal B}{\cal C}$ and
${\cal D}_-\equiv \biggl(a_1{\partial\over\partial t_2} -
a_2{\partial\over\partial t_1}\biggr){\cal A}{\cal B}{\cal C}$),
which implies that one will overconstrain the matrix $P$ (from the
first set of flow equations, one sees that ${\cal D}_-=0$, and hence
one gets from (\ref{eq:flow2edge}), eight equations in the six
parameters $P^i_j$). Hence, the uplift of the edge to a Spin(7) is
quite likely to be impossible.

\section*{Possibility of introducing Weierstrass elliptic Functions
in $Spin(7)$ eight-folds including an $S^1$}

We now give some very compelling evidence in support of the
possibility of inclusion of Weierstrass elliptic functions in those
$Spin(7)$ eight-folds which are of the type $X_7\times S^1$, $X_7$
being a seven fold of $SU(3)$ structure.

Seven-dimensional manifolds of $G_2$ holonomy or $SU(3)$ structure
are required for getting ${\cal N}=1$ supersymmetry in four
dimensions from the eleven dimensional $M$-theory. Similarly,
$Spin(7)$ eight-folds are required for getting ${\cal N}=1$
supersymmetry in three dimensions from (the eleven dimensional) $M$
theory. One could explore the option of getting compact $Spin(7)$
uplifts using $S^1$'s instead of intervals by using the same flow
equations as derived in this paper, but further demanding
periodicity w.r.t. $t_1$ and $t_2$, of the solutions.

Assuming the existence of $Spin(7)$ eight-folds of the type
$X_7\times S^1$, one could first argue the existence of a
$G_2$-structure by noting the existence of a singlet in the
decomposition under $G_2\subset Spin(7)$ of the ${\bf 8}$ in the
fundamental spinorial representation (${\bf 8}\rightarrow{\bf 7} +
{\bf 1}$). Further, assuming that  Majorana-Weyl spinors
($\xi=\xi^+\oplus\xi^-$, the $\pm$ signs referring to the
chiralities) on the $Spin(7)$ eight-fold are nowhere vanishing,
there is a further reduction of the structure group to $G_{2+}\cap
G_{2-}=SU(3)$, the two $G_2$'s corresponding to the two chiralities
of $\xi^\pm$.

Having established the connection between (the use of) $Spin(7)$
eight-folds and seven-folds with $SU(3)$-structure, let us now move
to the main theme of this section - the possibility of inclusion of
Weierstrass elliptic functions in seven-folds with $SU(3)$ structure
and thereby $Spin(7)$ eight-folds of the type $X_7\times S^1$.

Using the results of \cite{dp}, explicit metrics for seven-folds
with $SU(3)$ structure were obtained. The ``shape" deformation
functions ``$A(z,{\bar z}; v,{\bar v})$" and ``$B(z,{\bar z}; v,
{\bar v})$", as indicated in \cite{fkmr}, could also be related to
elliptic functions - the seven dimensional $SU(3)$ structure does
not impose too many constraints if one allows wrapped $M5$-branes in
the analysis. The following are some interesting connections between
some concepts, thereby motivating further the idea of having
singular uplifts to seven dimensions of the Iwasawa manifold,
involving elliptic functions:

\begin{itemize}
\item
The $D7$-brane metric (relevant to ``cosmic strings" in
\cite{greenetal}) is given by:
\begin{equation}
ds_{10}^2=ds_8^2 + \tau_2(z)|\eta(\tau(z))|^4|z|^{-{N\over6}}|dz|^2,
\end{equation}
where $\tau_2\equiv$Im$\tau$, $\tau=a+ie^{-\phi}$, $a\equiv$ axion
and $\phi\equiv$dilaton, $\eta\equiv$ Dedekind eta function (see
below), $N\equiv$ the number of $D7$-branes, and $z$ is the complex
coordinate transverse to the $D7$-brane. Hence, one has one explicit
example of a metric involving $\eta$, which is related to theta
functions, as indicated below.

\item
The Jacobi theta function \cite{ww} function defined for two complex
variables $z$ and $\tau$ where Im$\tau>0$:
$\vartheta(z:\tau)=\sum_{-\infty}^\infty e^{i\pi n^2\tau + 2\pi i n
z}$. The theta function is related to the Dedekind eta function via:
$\vartheta(0;\tau)={\eta^2({\tau+1\over2})\over\eta(\tau+1)}$.

\item The Weierstrass
elliptic function \cite{ww} is a doubly periodic function with
periods 2$\omega_1$ and $2\omega_2$ such that $\omega_1\omega_2$ is
not real:
$${\cal P}(z;\omega_1,\omega_2)={1\over z^2} +\sum_{m,n\in{\bf Z}\setminus
(0,0)}\biggl({1\over(z-2m\omega_1-2n\omega_2)^2} -
{1\over(2m\omega_1+2n\omega_2)^2}\biggr).$$ ${\cal P}$ satisfies the
following cubic equation:
$$\biggl({d{\cal P}(z;\tau)\over dz}\biggr)^2={\cal P}^3(z;\tau) - g_2{\cal P}
-g_3$$ where $g_2=60\sum_{m,n\in{\bf
Z}\setminus(0,0)}{1\over(2m\omega_1+ 2n\omega_2)^4},\
g_3=140\sum_{m,n\in{\bf Z}\setminus(0,0)}{1\over(2m\omega_1
+2n\omega_2)^6}$. This is an equation of a torus, which could be
related to the Riemann surface that is referred to in \cite{dp}. The
torus degenerates, i.e., becomes singular along the discriminant
locus given by: $\Delta=g_2^3-27g_3^2=0$. There is the following
relation between the discriminant locus and the Dedekind eta
function: $\Delta=(2\pi)^{12}\eta^{24}$. The following relations are
true: ${\cal P}(z;\omega_1=1,\omega_2=\tau)=-{d^2\over
dz^2}\vartheta_{11}(z;\tau) + $ constant
$=\pi^2\vartheta^2(0;\tau)\vartheta^2_{10}(0;\tau)
{\vartheta^2_{10}(z;\tau)\over\vartheta^2_{11}(z;\tau)}+e_2(\tau)$,
where $e_2$ is one of the three roots $e_{i=1,2,3}$, of the cubic
equation $4t^3-g_2t-g_3=0$, and ${\cal P}(\omega_1)=e_1,\ {\cal
P}(\omega_2)=e_2,\ {\cal P}(-\omega_1-\omega_2) =e_3$\footnote{The
$e_i$'s are given by:
$e_1(\tau)={\pi^2\over3}(\vartheta^4(0;\tau)+\vartheta^4_{01}(0;\tau))$,
$e_2(\tau)=-{\pi^2\over3}(\vartheta^4(0;\tau)+\vartheta^4(0;\tau))$
$e_3(\tau)={\pi^2\over3}(\vartheta^4_{10}(0;\tau)-\vartheta^4(0;\tau))$,
where the three other theta functions are defined as: $
\vartheta_{01}(z;\tau)=\vartheta(z+{1\over2};\tau)$
$\vartheta_{10}(z;\tau)=e^{i{\pi\over4}+i\pi
z}\vartheta(z+{\tau\over2};\tau)$
$\vartheta_{11}(z;\tau)=e^{i{\pi\over4} +
i\pi(z+{1\over2})}\vartheta(z + {\tau+1\over2};\tau)$.}

\item
Consider a holomorphic function $f(z)$ and $a,b\in{\bf R}$. Define
two operators $S_a$ and $T_b$ as follows:$ (S_a f)(z)=f(z+a),\ (T_b
f)(z)=e^{i\pi b^2\tau + 2i\pi b z}f(z + b\tau)$. Then $S,T$ and a
phase factor form the generators of the nilpotent Heisenberg group
central to the group-theoretic way of understanding the Iwasawa
manifold. If $U(\lambda\in{\bf C},a,b)\in H\equiv$Heisenberg group,
then
$$U(\lambda,a,b)f(z)
=\lambda (S_a\circ T_bf)(z)=\lambda e^{i\pi b^2\tau + 2i\pi b
z}f(z+b\tau+a),$$ and $U$ is referred to as the theta representation
of the Heisenberg group \cite{rf}.

\item
(Inverse) Elliptic functions, as shown in \cite{c}, naturally figure
in the hyper-K\"{a}hler metrics in twistor spaces obtained by
deformations of Atiyah-Hitchin spaces and Legendre transform. Lets
elaborate upon this a little.

Deformations of Atiyah-Hitchin manifolds (written as hypersurface in
${\bf C}^3$: $x^2+y^2z=1$) of the type $x^2z+(yz+a)^2=z^2+a^2$ have
been considered. The twistor three-folds are obtained as holomorphic
sections $\Gamma({\cal O}_{\bf CP^1}(4))$ with the following similar
equation:
$x^2(\zeta)\eta(\zeta)+(y(\zeta)\eta(\zeta)+p(\zeta))^2=\eta(\zeta)
+p^2(\zeta)$, where $\eta$ is $\Gamma({\cal O}_{\bf CP^1}(4))$ and
the deformation $p$ is $\Gamma({\cal O}_{\bf CP^1}(2))$, and $\zeta$
is a ${\bf CP}^1$-valued coordinate. The `reality involution':
${\bar\eta}^{(2m)}(\zeta)=(-)^m({\bar\zeta})^{(2m)}(-{1\over{\bar\zeta}})$,
where $\eta^{(2m)}(\zeta)$ is $\Gamma({\cal O}_{\bf CP^1}(2m))$,
implies that $\eta(\zeta)$ has five independent parameters, i.e.,
$\eta(\zeta)=z+v\zeta+w\zeta^2-{\bar v}\zeta^3+{\bar z}\zeta^4$
($z,v\in{\bf C}$ and $w\in{\bf R}$) and the deformation
$p(\zeta)=a+b\zeta-{\bar a}\zeta^2, a\in{\bf C}, b\in{\bf R}$. An
$SL(2,{\bf C})$ transformation:
$\zeta\rightarrow{a\zeta+b\over-b\zeta+{\bar a}}$, $|a|^2+|b|^2=1$
can be used to restrict $p(\zeta)=\tilde{b}\zeta$. For generic
values of the five real parameters in $\eta(\zeta)$, one gets eight
points on an elliptic curve, $\gamma^2=\eta(z\zeta)$, corresponding
to the roots of $\eta(\zeta) +p^2(\zeta)=0$ -  the divisor for four
of these eight should correspond to $\Gamma(L^m)$, where $L^m$ are
holomorphic line bundles over $T{\bf CP}^1$ with
$e^{-m\xi\over\zeta}$ ($\xi$ being a fiber coordinate) the
transition functions. The splitting of the eight roots into two
groups of four each is determined by the following condition:
$\biggl(\int_\alpha^\infty
+\int_\beta^\infty-\int_{-{1\over{\bar\alpha}}}^\infty -
\int_{-{1\over{\bar\beta}}}^\infty\biggr){d\zeta\over\sqrt{\eta(\zeta)}}=2$;
the roots of $\eta(\zeta)+p^2(\zeta)=0$ are
$\alpha,\beta,-{1\over{\bar\alpha}}, -{1\over{\bar\beta}}$. Now, if
$x_1,x_2$ are roots of $\eta(\zeta)+\tilde{b} \zeta=0$ {\it after}
the $SL(2,{\bf C})$ transformation:
$$\left(\begin{array}{c}
\eta(\zeta) \\
p(\zeta) \\
\end{array}\right)\stackrel{\zeta\rightarrow{a\zeta+b\over-{\bar b}\zeta+a}}
{\longrightarrow}\left(\begin{array}{c}
{r_1\zeta^3-r_2\zeta^2-r_1\zeta\over(-{\bar b}\zeta+{\bar a})^4}\\
{\tilde{b}(a\zeta+b)\over(-{\bar b}\zeta+{\bar a})}\\
\end{array}\right),$$ $r_1,r_2\in{\bf R}$,
then the aforementioned constraint can be rewritten in terms of
inverse elliptic functions:
$${\cal P}^{-1}\biggl(x_1-{r_2\over3r_1}\biggr)+{\cal P}^{-1}\biggl(x_2-
{r_2\over3r_1}\biggr)-{\cal P}^{-1}\biggl(-{1\over{\bar
x}_1}-{r_2\over3r_1} \biggr)-{\cal P}^{-1}\biggl(-{1\over{\bar
x}_2}-{r_2\over3r_1}\biggr) ={m\sqrt{r_1}\over2},$$ where the
inverse elliptic function ${\cal P}^{-1}(z)\equiv\int_\infty^z
{d\zeta\over\sqrt{4\zeta^3-g_2\zeta-g_3}}$. The constraint is also
equivalent to ${\partial F\over\partial w}=0$ for a suitable
constraint function $F$ defined in terms of appropriate contour
integrals. The K\"{a}hler potential is then given in terms of the
Legendre transform of $F: K(z,{\bar z};{\partial F\over\partial
v},{\partial{\bar F}\over
\partial{\bar v}})=F(z,{\bar z};v,{\bar v};w)-v{\partial F\over\partial v}
-{\bar v}{\partial{\bar F}\over\partial{\bar v}}$ evaluated at the
constraint: ${\partial F\over\partial w}=0$.

Thus, one sees the existence of (inverse) elliptic functions in the
hyper-K\"{a}hler metrics in twistor spaces obtained using
deformations of Atiyah-Hitchin spaces and Legendre transforms.
\end{itemize}

To summarize, we have obtained the relevant flow equations for
uplifting half-flat manifolds to Spin(7) eight-folds by two methods
- first, by considering a double $G_2$ (constructed from the
half-flat) -fibration over an interval, and the second, by
considering a fibration of the half-flat over the product of two
intervals. We were able to explicitly uplift the Iwasawa at the
standard complex structure limit in the moduli space of almost
complex structures on the Iwasawa. We gave a plausibility argument
against the same for the second method, at the ``edge". We also gave
motivating reasons for considering singular uplifts involving doubly
periodic functions - the physical interpretation of the same is not
yet clear.

\section*{Acknowledgement}

We wish to thank the Harvard theory group (S.Minwalla in particular)
and IAS, Princeton for the hospitality during the stay there where
this work was done, J.Maldacena and O.Lunin for discussions, and the
govt. of India for a research grant (under the Department of Atomic
Energy Young Scientist Award Scheme).


\begin{thebibliography}{99}
\bibitem{fkmr}A.~Franzen, P.~Kaura, A.~Misra and R.~Ray,
{\it Uplifting the Iwasawa}, arXiv:hep-th/0506224, to appear in
Fortschritte der Physik.
\bibitem{hitchin} N.~Hitchin, {\it Stable forms and special metrics}, in
{\it Global Differential Geometry: The mathematical legacy of Alfred
Gray}, 70, AMS, 2001 [arXiv:math.DG/0107101].
\bibitem{greenetal}B.~R.~Greene, A.~D.~Shapere, C.~Vafa and S.~T.~Yau,
{\it Stringy Cosmic Strings and Noncompact Calabi-Yau Manifolds},
Nucl.\ Phys.\ B {\bf 337}, 1 (1990).
\bibitem{c}G.~Chalmers,
{\it The implicit metric on a deformation of the Atiyah-Hitchin
manifold}, Phys.\ Rev.\ D {\bf 58}, 125011 (1998)
[arXiv:hep-th/9709082].
\bibitem{cardosoetal}G.~L.~Cardoso, G.~Curio, G.~Dall'Agata, D.~Lust,
P.~Manousselis and G.~Zoupanos, {\it Non-Kaehler string backgrounds
and their five torsion classes,} Nucl.\ Phys.\ B {\bf 652}, 5 (2003)
[arXiv:hep-th/0211118].
\bibitem{gurierietal}S.~Gurrieri, J.~Louis, A.~Micu and D.~Waldram,
{\it Mirror symmetry in generalized Calabi-Yau compactifications},
Nucl.\ Phys.\ B {\bf 654}, 61 (2003) [arXiv:hep-th/0211102].
\bibitem{dp}G.~Dall'Agata and N.~Prezas,
{\it N = 1 geometries for M-theory and type IIA strings with
fluxes,}
    Phys.\ Rev.\ D {\bf 69}, 066004 (2004)
      [arXiv:hep-th/0311146].
\bibitem{joyce}D.D.Joyce, {\it Compact Riemannian 7-Manifolds with Holonomy
$G_2$, I}, Jour. Diff. Geom. {\bf 43} (1996), 291; {\it Compact
Riemannian 7-Manifolds with Holonomy $G_2$, II}; Jour. Diff. Geom.
{\bf 43} (1996) 329; {\it Compact 8-Manifolds with Holonomy
$Spin(7)$}, Inv. Math. {\bf 123} (1996) 507.
\bibitem{Gubseretal} A. Brandhuber, J.Gomis S.S.Gubser and S.Gukov,
{\it Gauge Theory at large N and new $G_2$ Holonomy metrics},
arXiv:hep-th/0106034; M.Cvetic, G.W.Gibbons, H. Lu and C.N. Pope,
{\it New Complete non-compact $Spin(7)$ Manifolds}, hep-th/0103155;
{\it Supersymmetric $M$-branes and $G_2$ Manifolds},
arXiv:hep-th/0106126; {\it Resolved Branes and $M$-Theory on Special
Holonomy Spaces}, arXiv:hep-th/0106177.
\bibitem{ww}E.T.Whittaker and G.N.Watson, {\it A Course of Modern
Analysis}, Cambridge Univeristy Press (1952).
\bibitem{rf}H.E.Rauch and H.M.Farkas, {\it Theta Functions with Applications
to Riemann Surfaces}, (1974) Williams and Wilkins Co., Baltimore
(USA).
\end{thebibliography}
\end{document}